# Reliability Modeling of Single-Sided Aluminized Polyimide Films during Storage Considering Stress-Induced Degradation Mechanism Transition

Shi-Shun Chen, Dong-Hua Niu, Wen-Bin Chen, Jia-Yun Song, Ya-Fei Zhang, Xiao-Yang Li, *Member, IEEE* and Enrico Zio, *Fellow, IEEE*

*Abstract*—Single-sided aluminized polyimide films (SAPF) are widely used in thermal management of aerospace systems. Although the reliability of SAPF in space environments has been thoroughly studied, its reliability in ground environments during storage is always ignored, potentially leading to system failure. This paper aims to investigate the reliability of SAPF in storage environments, focusing on the effects of temperature and relative humidity. Firstly, the relationship between the performance degradation of SAPF and aluminum corrosion is identified. Next, considering the presence of two distinct stages in the influence of temperature on aluminum corrosion, a novel degradation model accounting for the degradation mechanism transition is developed. Additionally, a parameter analysis method is proposed for determining SAPF degradation mechanism based on experimental data. Then, a statistical analysis method incorporating an improved rime optimization algorithm is employed for parameter estimation, and the reliability model is established. Experimental results demonstrate that the proposed method effectively identifies two distinct stages in the impact of temperature on SAPF performance degradation. Furthermore, the proposed degradation model outperforms traditional degradation models with unchanged degradation mechanism in terms of degradation prediction accuracy, extrapolation capability and robustness, indicating its suitability for describing the degradation pattern of SAPFs.

## NOMENCLATURE

### A. List of Acronyms

| | |
|---|---|
| AIC | Akaike information criterion |
| BM | Brownian motion |
| CMD | Cramér-von Mises distance |
| KLD | Kullback-Leibler divergence |
| MC | Monte Carlo |
| RMSE | Root mean square error |
| SAR | Solar absorption ratio |
| SAPF | Single-sided aluminized polyimide film |
| TERIME | Improved RIME algorithm with enhanced exploration and exploitation |

### B. List of Symbols

| | |
|---|---|
| $B(t)$ | Brownian motion |
| $K$ | Total number of time points |
| $L$ | Lifetime |
| $L_{all}$ | Likelihood function given all degradation data |
| $L_{li}$ | Likelihood function given $\boldsymbol{y}_{li}$ |
| $L_{max}$ | Maximum value of the log-likelihood function |
| $L_q$ | Lifetime of the $q^{th}$ degradation trajectory |
| $N$ | Normal distribution |
| $P$ | Number of simulated degradation trajectories |
| Pr | Probability measure |
| $R$ | Reliability |
| $R_{all}$ | Predicted reliability distributions of the model fitted by all the degradation data |
| $R_{test}$ | Predicted reliability distributions of the model fitted by test degradation data |
| $T$ | Temperature |
| $T^*$ | Standardized temperature |
| $T_0$ | Lowest temperature level |
| $T_H$ | Highest temperature level |
| $T_{threshold}$ | Temperature threshold for degradation mechanism transition |
| $Y$ | SAR |
| $Y_0$ | Initial SAR |
| $f_{all}$ | Predicted lifetime distributions of the model fitted by all the degradation data |
| $f_{test}$ | Predicted lifetime distributions of the model fitted by test degradation data |
| $k$ | Number of stress levels in the degradation experiment |
| $m_{li}$ | Number of measurements for unit $i$ at the $l^{th}$ stress level |
| $n_l$ | Number of samples at the $l^{th}$ stress level |
| $n_p$ | Number of unknown model parameters |
| $r_2$ | Coefficient of determination |
| $s$ | Stress vector |
| $s_l$ | Stress vector of the $l^{th}$ stress level |
| $t$ | Storage time |
| $t_i$ | $i^{th}$ discrete time points |
| $\boldsymbol{t}_{li}$ | Measurement time vector of unit $i$ at the $l^{th}$ stress level |
| $t_{lij}$ | $j^{th}$ measurement time of unit $i$ at the $l^{th}$ stress level |
| $\boldsymbol{y}_{high}$ | Degradation data with temperature level at $T > T_{threshold}$ |
| $\boldsymbol{y}_{li}$ | Degradation data vector of unit $i$ at the $l^{th}$ stress level |
| $y_{lij}$ | jth degradation value of unit $i$ at the $l^{th}$ stress level |
| $\boldsymbol{y}_{low}$ | Degradation data with temperature level at $T < T_{threshold}$ |
| $\boldsymbol{\Xi}_{li}$ | Covariance matrix of the degradation data vector of unit $i$ at the $l^{th}$ stress level |
| $\alpha_1, \alpha_3$ | Parameters affecting the impact of temperature on degradation |
| $\alpha_2$ | Parameter affecting the impact of relative humidity on degradation |

This work was supported by the National Natural Science Foundation of China [grant number 51775020], the China Scholarship Council [grant number 202406020189], the Postdoctoral Fellowship Program of CPSF [grant number GZC20233365], the Fundamental Research Funds for Central Universities [grant number JKF-20240559], and the National Natural Science Foundation of China [grant numbers 62073009]. (*Corresponding author: Xiao-Yang Li.*)
Shi-Shun Chen is with the School of Reliability and Systems Engineering, Beihang University, Beijing, China, and with the Energy Department, Politecnico di Milano, Milan, Italy (e-mail: css1107@buaa.edu.cn).

Dong-Hua Niu, Jia-Yun Song and Ya-Fei Zhang are with the Beijing Tianyu Aerospace New Materials Technology Co., Ltd, Beijing, China.
Wen-Bin Chen and Xiao-Yang Li are with the School of Reliability and Systems Engineering, Beihang University, Beijing, China (e-mail: chenwenbin@buaa.edu.cn, leexy@buaa.edu.cn).
Enrico Zio is with the Energy Department, Politecnico di Milano, Milan, Italy, and with the Centre de Recherche sur les Risques et les Crises (CRC), MINES ParisPSL University, Sophia Antipolis, France (e-mail: enrico.zio@polimi.it).

| | |
|---|---|
| $\beta$ | Parameter affecting the time scale function |
| $\delta$ | Kronecker delta function |
| $\boldsymbol{\theta}$ | Unknown parameter vector |
| $\mu_{Y0}$ | Mean value of the initial SAR |
| $\mu_a$ | Mean value of $a$ |
| $\sigma$ | Diffusion coefficient |
| $\sigma_{Y0}$ | Standard deviation of the initial SAR |
| $\sigma_a$ | Standard deviation of $a$ |
| $\sigma_\varepsilon$ | Standard deviation of the measurement error |
| $\varphi_1(T)$ | Standardized function for temperature when $T < T_{threshold}$ |
| $\varphi_3(T)$ | Standardized function for temperature when $T > T_{threshold}$ |
| $\phi$ | Relative humidity |
| $\phi^*$ | Standardized relative humidity |
| $\phi_0$ | Lowest relative humidity level |
| $\phi_H$ | Highest relative humidity level |

## I. INTRODUCTION

### A. Background

Single-sided aluminized polyimide films (SAPFs) are advanced materials that play a vital role in the thermal management of aerospace systems, such as satellites, spacecraft and rockets [1, 2]. SAPFs can efficiently reflect solar and infrared radiation while providing electrical insulation, making them ideal as the outermost layer in systems requiring thermal stability [3]. The failure of SAPFs exposes critical components to harsh environments, which can ultimately result in overall system failure [4]. Therefore, accurate reliability prediction of SAPFs is essential to ensure system reliability. Since SAPF failure is mainly caused by performance degradation over time, it is crucial to develop a precise model that describes its degradation process.

SAPFs are primarily intended for space missions. Thus, several studies have been carried out to investigate their degradation mechanism in space, including the effects of electron exposure [5], atomic oxygen exposure [6], ultraviolet radiation [7] and proton irradiation [8]. These studies have laid the foundation for the reliable application of SAPFs in space. However, aerospace products also experience ground environments during storage before practical use [9]. Prolonged storage durations or harsh storage conditions can significantly degrade SAPF performance, potentially leading to the failure of aerospace mission. Despite this risk, the degradation of SAPFs during storage is ignored in existing studies. Since ground environments are highly distinct from space environments, it is crucial to investigate how the performance of SAPF degrades in response to storage time and varying storage conditions for ensuring long-term system reliability.

### B. Structure and Performance Degradation of SAPFs

To meet the strict weight requirement of aerospace applications, SAPFs are designed to be extremely thin, typically with a total thickness not exceeding 10 μm. The structure of a typical SAPF is illustrated in Fig. 1, with a 3.8 μm aluminum layer as the coating and a 6 μm polyimide layer as the substrate.

The solar absorption ratio (SAR), indicating the capacity of materials to absorb solar radiation, is a key parameter for evaluating the performance of SAPF in thermal management. When the aluminum coating of SAPF peels off, its SAR increases. During the storage in ground environments, temperature and relative humidity are the primary environmental factors influencing the peeling of the aluminum coating. Garnier et al. [10] used optical microscopy to observe the aluminum coating peeling process with the effect of hydrothermal aging, demonstrating that high temperature and humidity significantly accelerate the peeling process. Through chemical analysis and scanning electron microscopy, Zhou et al. [11] further demonstrated that the peeling of the aluminum coating was driven by aluminum corrosion rather than by substrate aging. The underlying mechanism of aluminum corrosion process involves the reaction of aluminum with moisture and oxygen, which can be represented by:

$$O_2 + H_2O + 4e^- \to 4OH^-, \ Al \to Al^{3+} + 3e^-, \\ Al^{3+} + 3OH^- \to Al(OH)_3. \quad (1)$$

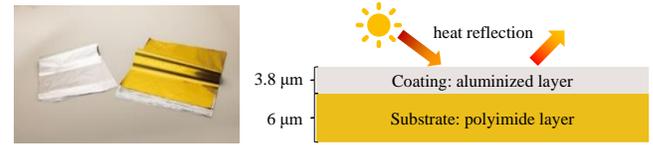

Fig. 1 The structure of the SAPF.

Once the corrosion of aluminum occurs, it is well-known that a dense $Al_2O_3$ layer is formed on its surface, which can described as:

$$2Al(OH)_3 \to Al_2O_3 + 3H_2O. \quad (2)$$

The $Al_2O_3$ layer can significantly reduce the corrosion rate to less than 1 μm per year in atmospheric conditions [12]. Although this corrosion rate may have a negligible impact on common aluminum alloys, it can critically affect ultra-thin SAPFs, where micron-scale corrosion significantly degrades performance. Therefore, the performance degradation of SAPF during storage should not be ignored.

In addition to $Al_2O_3$, existing studies have shown that $AlO(OH)$ is also formed when the temperature increases. Cao et al. [13] investigated aluminum alloy corrosion at six different temperature levels, and identified two distinct stages in the impact of temperature on aluminum corrosion process:

1) When the temperature is below around 60 ℃, only $Al_2O_3$ forms on the aluminum surface, as indicated by (2). In this stage, as the temperature increases, the corrosion rate accelerates.
2) When the temperature exceeds around 60 ℃, the aluminum surface forms not only $Al_2O_3$ but also $AlO(OH)$, as indicated by:

$$Al(OH)_3 \to AlO(OH) + H_2O. \quad (3)$$

The passivation film formed by $AlO(OH)$ is much denser than that of $Al_2O_3$. In this stage, as the temperature increases, the corrosion rate decreases.

Apart from the findings of Cao et al. [13], several other works have also described the same phenomenon that temperature affects aluminum corrosion in two distinct stages with $AlO(OH)$ formation at high temperature levels [14-16]. In this study, we refer to this behavior as stress-induced degradation mechanism transition, which is distinguished from multi-phase degradation that characterizes degradation transitions in the time dimension.

In practice, the possible ground storage environments of SAPFs vary greatly due to different usage requirements. In certain extreme areas, the land surface temperature can rise to 80 °C [17]. Consequently, it is necessary to consider the potential stress-induced degradation mechanism transition of SAPFs caused by temperature in degradation modeling for accurate reliability prediction.

*C. Related Works and Contributions*

Although several studies have demonstrated that temperature influences aluminum corrosion in two distinct stages, its impact on the performance degradation of SAPFs in thermal management remains unclear. From a system-level perspective, a reaction rate model for aluminum corrosion and oxide formation should be developed [13-16]. Then, by establishing the relationship between the aluminum content, aluminum oxide content and SAR, the performance degradation of SAPF in thermal management can be described. However, the aluminum oxide layer formed on the SAPF surface is extremely thin, typically around 5 nm [18]. Previous studies have shown that when the thickness is below 10 nm, it becomes challenging to determine and quantify the compound on the surface using techniques such as X-ray diffraction [19, 20]. Therefore, the aluminum corrosion process on the surface of SAPF is hard to be quantified, and the relationships between aluminum oxide content and SAR cannot be established.

In light of the above limitation, the SAPF must be treated as a "black box", with its degradation mechanism inferred from the SAR degradation data. To determine its degradation mechanism, a parameter analysis method is presented. Since SAR degradation is directly influenced by aluminum corrosion, the impact of temperature on SAR degradation may also exhibit two distinct stages. Therefore, two degradation models are developed to capture the influence of temperature at different stress levels according to the temperature threshold outlined in [13]. By analyzing the parameters related to temperature of two different degradation models, the degradation mechanism of SAPF can be determined.

However, there remains a critical research gap in degradation modeling of SAPFs. Current studies primarily focus on degradation modeling using stochastic processes [21-24], such as Wiener process [25, 26], Gamma process [27, 28] and inverse Gaussian process [29, 30], under constant stress conditions. However, in practical applications, the storage conditions of SAPF vary continuously in response to product demand, leading to corresponding variations in the degradation rate. Therefore, it is essential to incorporate the effects of changing storage conditions into the degradation model. When considering the influence of stress, an unchanged degradation mechanism is always applied in existing studies to model the degradation [31-36], which indicates that the impact of a stress on the degradation rate across all stress levels follows a unified stress model, like the Arrhenius model for temperature, the Power law model for humidity, and the exponential law model for electrical stress [37]. Nonetheless, SAPF degradation involves potential mechanism transitions, requiring distinct models for different degradation mechanisms. Although changes of the degradation mechanism in the time dimension have been studied and multi-phase degradation models have been proposed [38-42], these models will produce biased reliability predictions for SAPFs, as the mechanism transition in SAPFs is stress-induced rather than time-driven. As a result, existing methods may not be suitable for SAPF degradation modeling.

To address this issue, a novel degradation model considering the stress-induced degradation mechanism transition is constructed to describe the degradation patterns of SAPFs across all temperature levels. Besides, multi-source uncertainties are considered in the degradation model, and the most suitable model for quantifying the uncertainties in the degradation process is identified by comparative studies.

The main contributions of this study are summarized as follows:
1) A storage reliability prediction model of SAPF is established with the influences of temperature and relative humidity, where the degradation mechanism transition caused by temperature and multi-source uncertainties in the degradation process are considered.
2) A parameter analysis method is presented to determine the degradation mechanisms of SAPF with temperature based on SAR degradation data.
3) The effectiveness of the proposed method in describing the degradation pattern of SAPFs is verified by comparative studies, and the most suitable model for quantifying the uncertainties in the degradation process is identified.

It is worth noting that although the SAPF is the primary motivation and application scenario for the proposed method, the approach is broadly applicable to other products that exhibit stress-induced degradation mechanism transitions. Moreover, to the best of our knowledge, this is the first study to model the degradation and reliability of products exhibiting stress-induced degradation mechanism transitions, thereby extending the applicability of existing reliability modeling methodologies.

The organization of the paper is as follows. The proposed reliability model considering stress-induced degradation mechanism transition and multi-source uncertainties are presented in Section II, as well as the degradation mechanism determination method. Next, the degradation modeling results and analysis are provided in Section 0. After that, comparative studies and reliability predictions are conducted in Section IV. Finally, Section VI concludes the work.

## II. Proposed Method

*A. Model Construction*

For the SAR degradation considering the impacts of temperature ($T$) and relative humidity ($\phi$), its deterministic performance degradation can be described as [32]:

$$Y(t,s) = Y_0 + ae(s)\Lambda(t), \quad (4)$$

where $Y$ represents the SAR; $Y_0$ is the initial SAR without considering degradation; $t$ is the storage time; $\Lambda(\cdot)$ is the timescale function and $\Lambda(t) = t^\beta$ is used for degradation trajectory description in this study, capable of representing both linear and nonlinear degradation processes [43]; $a$ is an unknown parameter; $e(s)$ is the stress-related degradation rate model given by:

$$e(s) = \exp(\alpha_1 T^* + \alpha_2 \phi^*), s = [T, \phi], \quad (5)$$

where $\alpha_1$ and $\alpha_2$ are unknown parameters; $T^*$ and $\phi^*$ are the standardized stress calculated by the Arrhenius model and the Power law model, respectively, as [44]:

$$T^* = \frac{1/T_0 - 1/T}{1/T_0 - 1/T_H}, \quad \phi^* = \frac{\ln\phi - \ln\phi_0}{\ln\phi_H - \ln\phi_0}, \quad (6)$$

where $T_0$ and $T_H$ denote the lowest and highest stress levels of the temperature, respectively; $\phi_0$ and $\phi_H$ denote the lowest and highest stress levels of the relative humidity, respectively.

A general degradation process with unchanged degradation mechanism can be effectively described by (4) - (6). However, based on the analysis in Section I.B, the influence of temperature on SAPF performance degradation may exhibit two distinct stages, determined by the temperature stress level and the corresponding threshold. To describe the degradation mechanism transition, (5) is expended as:

$$e(s) = \begin{cases} \exp(\alpha_1 \varphi_1(T) + \alpha_2 \phi^*), & T < T_{threshold} \\ \exp(\alpha_1 + \alpha_2 \phi^* + \alpha_3 \varphi_3(T)), & T \geq T_{threshold} \end{cases}, \quad (7)$$

where $T_{threshold}$ represents the temperature threshold for the degradation mechanism transition. Inspired by (6), $\varphi_1(T)$ and $\varphi_3(T)$ are defined by:

$$\varphi_1(T) = \frac{1/T_0 - 1/T}{1/T_0 - 1/T_{threshold}}, \varphi_3(T) = \frac{1/T_{threshold} - 1/T}{1/T_{threshold} - 1/T_H}. \quad (8)$$

For the degradation models (4), (7) and (8), several characteristics can be deduced:
1) Regardless of the relationship between $T$ and $T_{threshold}$, the values of $Y_0$, $a$, $\alpha_1$, $\alpha_2$, and $\beta$ remain unchanged. The value of $\alpha_2$ is a constant, indicating that the influence mechanism of humidity on SAPF degradation remains unchanged.
2) When $T < T_{threshold}$, the impact of temperature on SAPF degradation is governed by $\alpha_1$, which is related to the aluminum corrosion process described by (2) at low temperature levels; when $T \geq T_{threshold}$, it is controlled by $\alpha_3$, which is related to the aluminum corrosion process described by (2) and (3) at high temperature levels. The distinct parameters in each stage reflect the varying influence mechanism of temperature on SAPF degradation.
3) As $T$ approaches $T_{threshold}$, $\varphi_1(T)$ approaches unity, and $\varphi_3(T)$ approaches zero. At this point, the degradation models for both stages ultimately converge to:

$$Y(t,\phi) = Y_0 + a \exp(\alpha_1 + \alpha_2 \phi^*) t^\beta. \quad (9)$$

Equation (4) describes the deterministic performance degradation of the SAPF. However, in practice, the degradation processes are always influenced by various uncertainties, and quantifying these uncertainties is crucial for accurate reliability prediction. Firstly, variations in production and manufacturing processes introduce uncertainties in the physical properties of SAPFs. These variations lead to uncertainties in both the initial performance and the degradation rate of SAR. Besides, due to the difficulty of measuring the optical parameter SAR and the requirement for manual positioning of SAPFs, measurement errors are inevitable. Furthermore, the dispersion of the performance degradation values always increases over time due to uncertainty in the time dimension. This phenomenon is generally described by Brownian motion (BM), also known as the Wiener process. The uncertainties mentioned above have been extensively studied in degradation modeling, and mature quantification methods have been established [32, 45, 46].

Based on the above analysis, the constructed degradation model, which accounts for the influences of $T$ and $\phi$, stress-induced degradation mechanism transition and multi-source uncertainties, is presented by:

$$Y(t,s) = Y_0 + ae(s)t^\beta + \sigma B(t) + \varepsilon,$$

$$e(s) = \begin{cases} \exp(\alpha_1 \varphi_1(T) + \alpha_2 \phi^*), & T < T_{threshold} \\ \exp(\alpha_1 + \alpha_2 \phi^* + \alpha_3 \varphi_3(T)), & T \geq T_{threshold} \end{cases}, \quad (10)$$

$$Y_0 \sim N(\mu_{Y_0}, \sigma_{Y_0}^2), a \sim N(\mu_a, \sigma_a^2), \varepsilon \sim N(0, \sigma_\varepsilon^2),$$

where $\mu_i$ and $\sigma_i$ represent the mean and standard deviation of the random variable $i$, respectively; $\sigma$ is the diffusion coefficient; $\varepsilon$ represents the measurement error; $B(t)$ denotes a standard BM; and $N$ represents the normal distribution.

**Remark 1**: While multiple sources of uncertainties are considered in (10), not all of them significantly impact actual SAPF degradation. Comprehensive quantification of these uncertainties enables development of targeted strategies to improve SAPF reliability. For example, uncertainties related to performance initial values and degradation rate can be reduced through manufacturing process enhancements. Additionally, by comparing different ablation model variants, as discussed in Section 4.1, the most suitable degradation model for SAPF can be identified.

*B. Statistical Analysis*

According to (8) and (10), the unknown parameters to be determined from the observed degradation data are $\boldsymbol{\theta} = \left[\mu_{Y_0}, \sigma_{Y_0}, \mu_a, \sigma_a, \alpha_1, \alpha_2, \alpha_3, \beta, \sigma, \sigma_\varepsilon, T_{threshold}\right]$. To ensure the generality of the proposed statistical analysis method, it is assumed that there are $k$ stress levels in the constant-stress degradation experiments, with $n_l$ samples at the $l^{th}$ stress level. Then, the $j^{th}$ degradation value of unit $i$ at the $l^{th}$ stress level is denoted as $y_{lij}$, where $j = 1, 2,..., m_{li}$, and $m_{li}$ represents the number of measurements for unit $i$ at the $l^{th}$ stress level. The corresponding measurement time is indicated by $t_{lij}$. Hereby, we denote $\boldsymbol{y}_{li} = \left[y_{li1}, y_{li2}, ..., y_{lim_{li}}\right]^T$ and $\boldsymbol{t}_{li} = \left[t_{li1}^\beta, t_{li2}^\beta, \cdots, t_{lim_{li}}^\beta\right]^T$. Then, according to the property of the standard BM, $\boldsymbol{y}_{li}$ follows a multivariate normal probability distribution given by:

$$\boldsymbol{y}_{li} \sim N\left(\mu_{Y_0} + \mu_a e(s_l)\boldsymbol{t}_{li}, \boldsymbol{\Xi}_{li}\right), \quad (11)$$

and $\boldsymbol{\Xi}_{li}$ is a $m_{li} \times m_{li}$ dimensional covariance matrix with its $(u, v)^{th}$ entry calculated by:

$$(\boldsymbol{\Xi}_{li})_{uv} = \sigma_{Y_0}^2 + \sigma_a^2 e^2(s_l) t_{liu}^\beta t_{liv}^\beta + \sigma^2 \min(t_{liu}, t_{liv}) + \delta_{uv}\sigma_\varepsilon^2, \quad (12)$$

where $\delta_{uv}$ is the Kronecker delta function. When $u = v$, $\delta_{uv} = 1$; otherwise, $\delta_{uv} = 0$.

Then, the log-likelihood function given $\boldsymbol{y}_{li}$ can be deduced as:

$$\ln L_{li}(\boldsymbol{\theta}|\boldsymbol{y}_{li}) = -\frac{1}{2}\Big[m_{li}\ln(2\pi) + \ln|\boldsymbol{\Xi}_{li}|$$
$$+ (\boldsymbol{y}_{li} - \mu_{Y_0} - \mu_a e(s_l)\boldsymbol{t}_{li})^T \boldsymbol{\Xi}_{li}^{-1}(\boldsymbol{y}_{li} - \mu_{Y_0} - \mu_a e(s_l)\boldsymbol{t}_{li})\Big]. \quad (13)$$

Based on (13), the log-likelihood function given all the observations $\boldsymbol{y}$ can be expressed as:

$$\ln L_{all}(\boldsymbol{\theta}|\boldsymbol{y}) = \sum_{l=1}^{k}\sum_{i=1}^{n_l} \ln L_{li}(\boldsymbol{\theta}|\boldsymbol{y}_{li}). \quad (14)$$

| **Algorithm 1**: The statistical analysis method using TERIME to obtain the unknown parameters based on experiment data |
|---|
| **Input**: |
| 1. Degradation observation data $y$ and corresponding measurement time $t$ under all the stress levels. |
| 2. Lower and Upper boundaries of the unknown parameters. |
| 3. Population size and maximum number of iterations for TERIME. |
| **Output**: |
| 1. Estimates of the unknown parameters in the proposed degradation model. |
| 2. Maximum value of the log-likelihood function. |
| **Objective function**: |
| 1. Maximizing the log-likelihood function (14). |
| **Procedure**: |
| 1. Detailed steps of TERIME can be referred to [47]. |

By maximizing the log-likelihood function (14), the estimates of $\theta$ can be obtained. However, due to the complexity introduced by the multiple unknown parameters in the proposed degradation model, methods like the Nelder-Mead and Newton-Raphson approaches face challenges in finding the global optima. Therefore, an improved meta-heuristic rime optimization algorithm called TERIME [47] is employed due to its robust performance in high-dimensional optimization problems. Based on the conclusions in [47], the population size of TERIME is set to 20, and the maximum number of iterations is set to 50000. The details of the parameter estimation procedure are summarized in Algorithm 1.

### C. SAPF Degradation Mechanism Determination

Since the impact of temperature on the thermal management performance degradation of SAPF remains unclear, a degradation mechanism determination method is proposed in this section to specify the degradation mechanism of SAPFs described by (7).

According to the findings on aluminum corrosion in [13], the value of $T_{threshold}$ is approximately 60 ℃. Herein, we denote the degradation data at temperature levels $\geq 60\,°\mathrm{C}$ as $y_{high}$, and those $< 60\,℃$ as $y_{low}$. Then, a degradation model assuming an unchanged degradation mechanism with temperature is employed for degradation mechanism determination, which is expressed by:

$$Y(t,s) = Y_0 + e(s)t^\beta + \sigma B(t) + \varepsilon,$$
$$e(s) = a \exp\left(\alpha_1 \varphi_1 T^* + \alpha_2 \phi^*\right), \quad (15)$$
$$Y_0 \sim N\left(\mu_{Y_0}, \sigma_{Y_0}^2\right), a \sim N\left(\mu_a, \sigma_a^2\right), \varepsilon \sim N\left(0, \sigma_\varepsilon^2\right).$$

Based on the aluminum corrosion mechanism described in Section I-B, if $\alpha_1 > 0$ for $y_{low}$ and $\alpha_1 < 0$ for $y_{high}$, it demonstrates that the influence of $T$ on SAPF performance degradation presents two distinct stages similar to aluminum corrosion. Otherwise, if $\alpha_1 > 0$ for both $y_{low}$ and $y_{high}$, the influence of $T$ on SAPF performance degradation is different from aluminum corrosion and is similar to general products.

The point estimates of $\alpha_1$ in (15) for $y_{low}$ and $y_{high}$ can be estimated according to Algorithm 1. Furthermore, considering the uncertainty in parameter estimation caused by potential

| **Algorithm 2**: The MC method to calculate the reliability |
|---|
| 1. Determine the unknown model parameters $\hat{\theta} = \left[\hat{\mu}_{Y_0}, \hat{\sigma}_{Y_0}, \hat{\mu}_a, \hat{\sigma}_a, \hat{\alpha}_1, \hat{\alpha}_2, \hat{\alpha}_3, \hat{\beta}, \hat{\sigma}, \hat{\sigma}_\varepsilon, \hat{T}_{threshold}\right]$ via the statistical analysis method in Section II.B. |
| 2. Specify the storage conditions $T$ and $\phi$. |
| 3. Simulate $B(t)$ based on the property of the standard BM. |
| 4. Sample a unit-specific $Y_0$ and $a$ from their distributions in (10), and acquire the degradation trajectory by (8) and (10). |
| 5. Repeat steps 3 and 4 until $P$ degradation trajectories are generated. |
| 6. Specify the performance requirement $Y_{th}$. |
| 7. Calculate the lifetime of the $P$ simulated degradation trajectories by (16). For the $q^{th}$ degradation trajectory, its lifetime is denoted as $L_q$. |
| 8. Obtain the empirical lifetime distribution of the product as $F\left(t\middle\vert\hat{\theta}\right) = \dfrac{1}{P}\sum_{q=1}^{P} 1_t(L_q)$, where $1_t(L_q) = \begin{cases} 1, L_q \leq t \\ 0, L_q > t \end{cases}$. |
| 9. Compute product reliability at a specific time $t$ by (17). |

insufficient sample size, interval estimates are also presented. While the non-parametric bootstrap is a commonly used technique for interval estimation in degradation models [32], it is not suitable here due to non-negligible measurement error in the SAPF degradation data as shown in Fig. 2, where resampling noisy samples tends to yield excessively wide intervals [48]. Therefore, subsampling (resampling without replacement) is adopted as an alternative [49]. Specifically, for each stress level, eight degradation paths are selected based on the recommended 63.2% subsampling ratio [50, 51], and parameter estimation is performed using Algorithm 1. Then, by repeating this procedure $W$ times, an empirical distribution of the parameter estimates is obtained, from which the interval estimates can be derived at a specific confidence level. The statistical significance of the sign of $\alpha_1$ is further evaluated by identifying the confidence level at which its confidence interval remains strictly above or below zero.

### D. Reliability Prediction

To ensure the thermal management performance of SAPF, its SAR is subject to a predefined requirement $Y_{th}$. As the storage duration extends, the SAR progressively increases. When the SAR exceeds $Y_{th}$, the SAPF is considered to be failed. According to the definition of the first passage time of a standard BM, the lifetime $L$ of SAPF can be calculated by:

$$L(s) = \inf\left\{t > 0 \middle\vert Y(t,s) \geq Y_{th}\right\}. \quad (16)$$

Then, the reliability of SAPF can be calculated as:

$$R(t,s) = \Pr\left\{L(s) \geq t\right\}, \quad (17)$$

where Pr represents the probability measure.

Due to the nonlinearity caused by the timescale function in the degradation model, it is challenging to derive closed-form analytical solutions for (16) and (17). Therefore, the Monte Carlo (MC) method is employed to calculate the lifetime distribution and reliability. The detailed steps of the calculation are provided in Algorithm 2. Moreover, using the $W$ sets of parameter estimates obtained from the subsampling in Section

TABLE I
PARAMETER ESTIMATION RESULTS OF THE MODEL WITH AN UNCHANGED DEGRADATION MECHANISM UNDER DIFFERENT TEMPERATURE LEVELS

| Parameters | Lower bound | Upper bound | Low temperature level | High temperature level |
|---|---|---|---|---|
| $\mu_{Y0}$ | 8 | 10 | 8.797, [8.747, 8.839] | 8.883, [8.836, 8.914] |
| $\sigma_{Y0}$ | 0 | 1 | 0.1693, [0.0930, 0.2023] | 0.1788, [0.1348, 0.2074] |
| $\mu_a$ | 0 | 10 | 0.01187, [0.00374, 0.03147] | 0.03316, [0.00528, 0.05006] |
| $\sigma_a$ | 0 | 1 | 5.25e-11, [2.56e-11, 1.1e-3] | 3.97e-10, [2.18e-11, 9.52e-10] |
| $\alpha_1$ | -10 | 10 | 2.109, [0.1337, 3.918] | -0.5817, [-1.158, 0.2616] |
| $\alpha_2$ | -10 | 10 | 0.8658, [0.2563, 1.600] | 0.1285, [0.073, 2.029] |
| $\beta$ | 0 | 2 | 0.4002, [0.3216, 0.4815] | 0.6745, [0.4715, 0.8120] |
| $\sigma$ | 0 | 1.5 | 8.19e-10, [8.47e-11, 1.44e-9] | 1.26e-9, [1.53e-110, 1.60e-9] |
| $\sigma_\varepsilon$ | 0 | 1 | 0.3576, [0.3375, 0.3766] | 0.3661, [0.3444, 0.3973] |

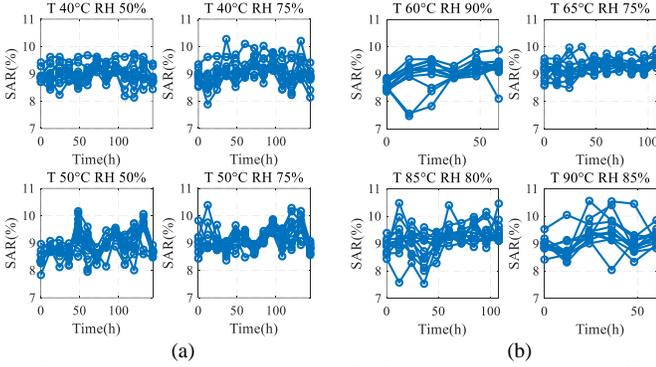

Fig. 2 Experimental degradation data of SAPF: (a) temperature levels < 60 °C; (b) temperature levels ≥ 60 °C.

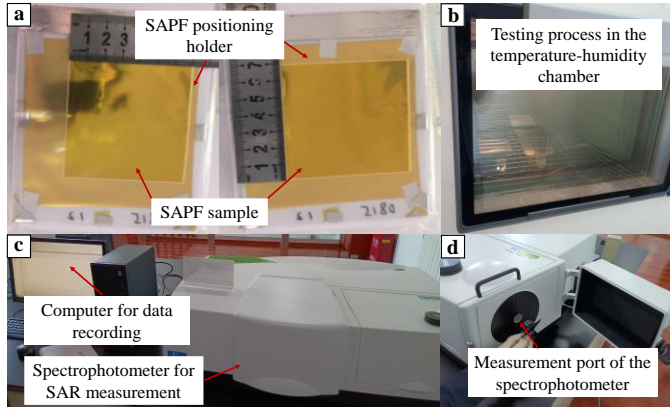

Fig. 3 Details of the SAPF degradation experiment: (a) the SAPF sample and the positioning holder (b) experiment process in the temperature-humidity chamber (c, d) the spectrophotometer for SAR measurement and the computer for data recording.

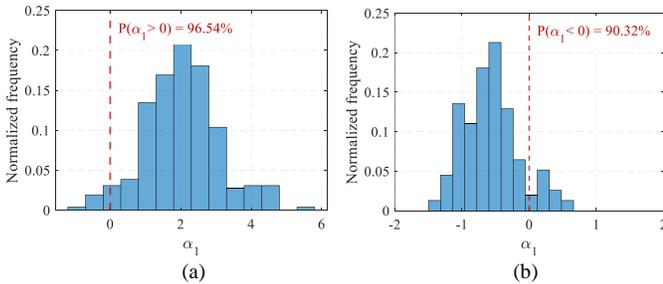

Fig. 4 Distribution of $\alpha_1$ from the interval estimation of the model with unchanged degradation mechanism: (a) at low temperature levels; (b) at high temperature levels.

II-C, the empirical reliability distribution at a specified time can be derived, which can then be used to establish a confidence interval of reliability at a certain confidence level.

## III. DEGRADATION MODELING RESULTS

### A. Data Description

In this study, we have SAPF degradation data at eight temperature and relative humidity levels. The stress levels and recorded data are presented in Fig. 2. The temperature and relative humidity are controlled by a temperature-humidity chamber during the testing. For each stress level, twelve SAPF samples were randomly selected from three different production batches. The SAPFs in the chamber were taken out every twelve hours to measure their SAR. To simulate the real storage state of SAPF and facilitate its fixation in the chamber, a positioning holder of polyethylene was designed, consisting of two clamping surfaces with a 7 cm ×7 cm square hole in the middle for performance measurement. A Lambda 1050S spectrophotometer manufactured by PerkinElmer was employed for SAR measurement. The samples and devices used in the degradation testing are illustrated in Fig. 3.

**Remark 2**: Ideally, to accurately determine the degradation mechanism, experiments should be conducted by first maintaining a constant $\phi$, then observing the SAR degradation under different $T$, and confirming the degradation mechanism with $T$. However, this approach will significantly increase both the experimental duration and observation costs, as it must be followed by additional tests under varying $\phi$ to assess its influence. Due to the financial constraints of the authors, this study is limited in its ability to perform such comprehensive experiments or extend testing across more stress levels. To address the limitations of the available data, a parameter analysis method is presented in this study, which satisfies the requirement for determining degradation mechanism with $T$. Besides, uncertainty analysis of the parameter estimates is performed to ensure result robustness and statistical significance. The proposed method is both effective and applicable in practical applications, allowing for degradation mechanism determination without requiring rigorous experimental design.

### B. Degradation Mechanism Determination

Based on the experimental degradation data illustrated in Fig. 2, the unknown parameters in (15) for $y_{low}$ and $y_{high}$ can be estimated according to Algorithm 1. Table I lists the parameter estimates and 90% confidence intervals for both $y_{low}$ and $y_{high}$. The repeated times $W$ for subsampling is set as 1000. From Table I, the point estimates show $\alpha_1 > 0$ for $y_{low}$ and $\alpha_1 < 0$ for $y_{high}$, suggesting that SAR degradation exhibits two distinct stages similar to the corrosion behavior of bulk aluminum. Additionally, Fig. 4 illustrates the empirical distribution of $\alpha_1$ obtained from the subsampling-based interval estimation. The value of $\alpha_1$ is positive in 96.54% of the estimates at the low temperature stress level, and negative in 90.32% at the high temperature stress level. These results indicate that the one-sided 90% confidence interval of $\alpha_1$ does not include zero, confirming that its sign is statistically significant across subsamples with a confidence level exceeding 90%. This provides strong evidence for a two-phase degradation mechanism driven by temperature in SAPFs. Therefore, it is necessary to employ the proposed degradation model considering the stress-induced degradation mechanism transition for accurately describing the degradation behavior of SAPFs.

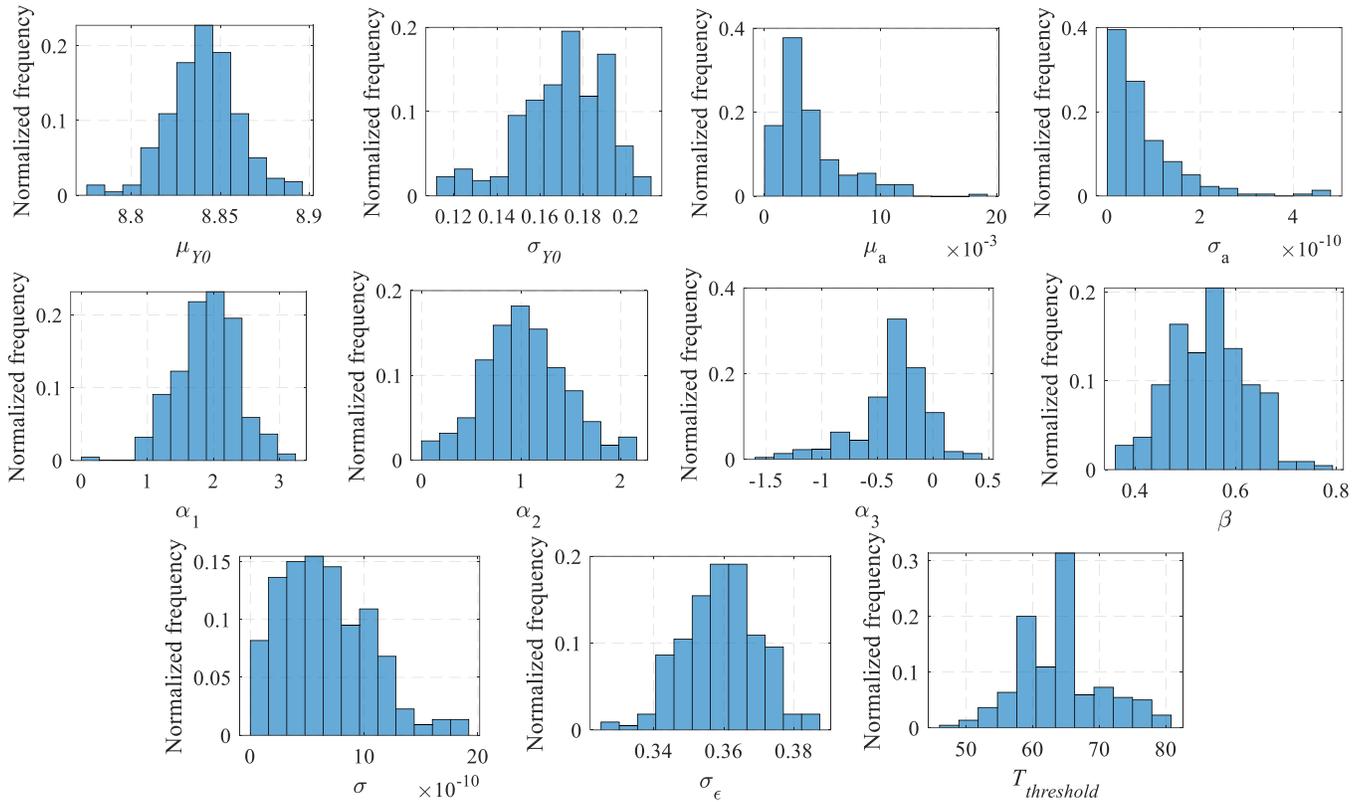

Fig. 5 Parameter distributions from the interval estimation of the proposed model using all data.

TABLE II
PARAMETER ESTIMATION RESULTS OF THE PROPOSED MODEL USING TERIME

| Parameters | Lower bound | Upper bound | Mean estimated values | Interval estimates |
|---|---|---|---|---|
| $\mu_{Y0}$ | 8 | 10 | 8.844 | [8.808, 8.873] |
| $\sigma_{Y0}$ | 0 | 1 | 0.1754 | [0.1261, 0.1986] |
| $\mu_a$ | 0 | 10 | 0.003374 | [0.000930, 0.01065] |
| $\sigma_a$ | 0 | 1 | 9.2416e-11 | [6.161e-12, 2.391e-10] |
| $\alpha_1$ | -10 | 10 | 1.836 | [1.107, 2.632] |
| $\alpha_2$ | -10 | 10 | 0.9439 | [0.3549, 1.756] |
| $\alpha_3$ | -10 | 10 | -0.2901 | [-0.9664, 0.05202] |
| $\beta$ | 0 | 2 | 0.5510 | [0.4204, 0.6715] |
| $\sigma$ | 0 | 1.5 | 5.833e-10 | [7.080e-11, 1.309e-9] |
| $\sigma_\varepsilon$ | 0 | 1 | 0.3609 | [0.3426, 0.3762] |
| $T_{threshold}$ | 40 | 80 | 63.75 | [54.30, 75.76] |

**Remark 3**: Analyzing the aluminum oxides on the SAPF surface at different temperature levels is an effective way to confirm the SAPF degradation mechanism. However, as mentioned in the introduction, due to the limited thickness of the aluminum oxide layer formed on the SAPF surface, which is below the analysis limits of current instruments, techniques such as X-ray diffraction are unable to determine and quantify the aluminum oxides on the surface.

### C. Degradation Modeling Results and Analysis

Based on the experimental degradation data reported in Fig. 2, the unknown parameters in the proposed degradation model (10) can be estimated according to Algorithm 1. Subsampling introduced in Section III-B is also employed for interval estimation to check the estimation robustness. The estimation results with intervals at 90% confidence level are listed in Table II, and the parameter distributions from interval estimations are illustrated in Fig. 5

From Table II, it can be seen that point estimate of $T_{threshold}$ is 63.75 ℃, aligning closely with the experimental threshold observed by Cao et al. [13]. Additionally, $\alpha_1 > 0$ and $\alpha_3 < 0$ indicate that the effect of temperature on SAR performance degradation exhibits a mechanism shift from accelerating degradation to inhibiting it, which is also consistent with the aluminum corrosion findings experimentally observed by Cao et al. [13]. The fact that $\alpha_1 > \alpha_2$ suggests that the effect of $T$ on performance degradation outweighs that of $\phi$ when $T < T_{threshold}$.

TABLE III
PARAMETER ESTIMATION RESULTS OF THE PROPOSED MODEL UNDER 20 INDEPENDENT RUNS USING TERIME

| Parameters | Lower bound | Upper bound | Mean estimated values | Standard deviation | Coefficient of variation |
|---|---|---|---|---|---|
| $\mu_{Y0}$ | 8 | 10 | 8.844 | 3.365e-9 | 3.805e-10 |
| $\sigma_{Y0}$ | 0 | 1 | 0.1754 | 2.498e-9 | 1.424e-8 |
| $\mu_a$ | 0 | 10 | 0.003374 | 3.708e-10 | 1.099e-7 |
| $\sigma_a$ | 0 | 1 | 9.242e-11 | 5.316e-11 | 0.575 |
| $\alpha_1$ | -10 | 10 | 1.836 | 7.294e-8 | 3.973e-8 |
| $\alpha_2$ | -10 | 10 | 0.9439 | 7.891e-8 | 8.359e-8 |
| $\alpha_3$ | -10 | 10 | -0.2901 | 4.425e-8 | -1.525e-7 |
| $\beta$ | 0 | 2 | 0.5510 | 1.116e-8 | 2.026e-8 |
| $\sigma$ | 0 | 1.5 | 5.833e-10 | 2.560e-10 | 0.439 |
| $\sigma_\varepsilon$ | 0 | 1 | 0.3609 | 1.237e-9 | 3.428e-9 |
| $T_{threshold}$ | 40 | 80 | 63.75 | 8.654e-7 | 1.358e-8 |

TABLE IV
PARAMETER ESTIMATION RESULTS OF THE PROPOSED MODEL UNDER 20 INDEPENDENT RUNS USING NELDER-MEAD

| Parameters | Lower bound | Upper bound | Mean estimated values | Standard deviation | Coefficient of variation |
|---|---|---|---|---|---|
| $\mu_{Y0}$ | 8 | 10 | 8.953 | 0.08398 | 0.00938 |
| $\sigma_{Y0}$ | 0 | 1 | 0.1857 | 0.02933 | 0.1579 |
| $\mu_a$ | 0 | 10 | 3.871 | 3.985 | 1.030 |
| $\sigma_a$ | 0 | 1 | 0.2420 | 0.2999 | 1.239 |
| $\alpha_1$ | -10 | 10 | -5.600 | 5.205 | -0.9293 |
| $\alpha_2$ | -10 | 10 | -1.308 | 5.490 | -4.197 |
| $\alpha_3$ | -10 | 10 | 2.0942 | 4.555 | 2.175 |
| $\beta$ | 0 | 2 | 0.6345 | 0.3139 | 0.4948 |
| $\sigma$ | 0 | 1.5 | 0.007694 | 0.006408 | 0.8329 |
| $\sigma_\varepsilon$ | 0 | 1 | 0.3679 | 0.006978 | 0.01897 |
| $T_{threshold}$ | 40 | 80 | 47.98 | 12.55 | 0.2617 |

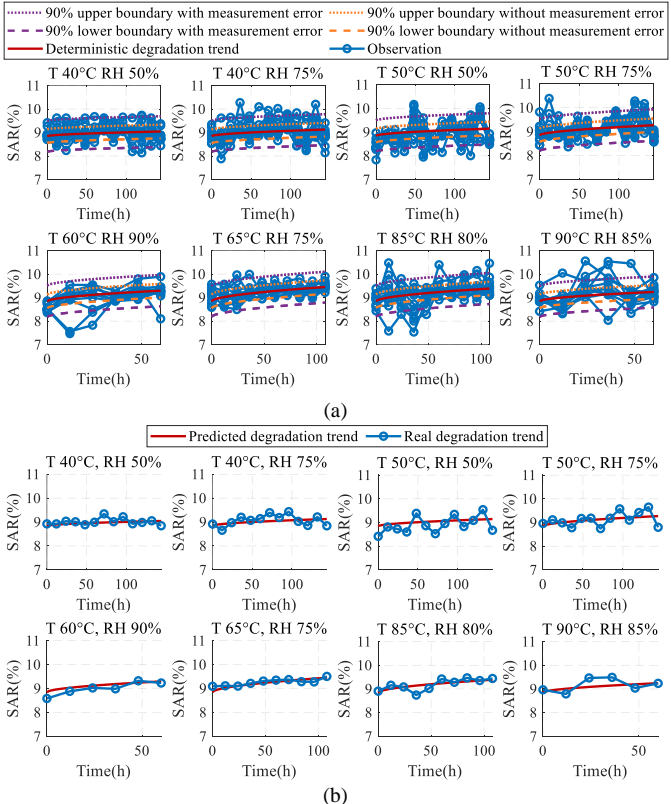

Fig. 6 Degradation predictions of SAR under all stress levels: (a) deterministic degradation trends with boundaries and observed degradation data (b) deterministic degradation trends and real degradation trends.

In terms of uncertainties, the fact that $\sigma_a$ is significantly smaller than $\mu_a$ suggests high consistency in the performance degradation of SAPF. This finding aligns with the degradation mechanism of SAPF. During storage, the performance degradation of SAPF is primarily governed by aluminum corrosion. Depending on the temperature level, a dense layer of $Al_2O_3$ or AlO(OH) forms on the aluminum surface. These passivation layers significantly slow down interfacial reactions, leading corrosion into a consistent process mainly influenced by environmental conditions, rather than by differences in microstructure or manufacturing. For the initial performance, $\sigma_{Y0}$ is 1.98% of $\mu_{Y0}$, which is within the expected quality control requirement of 3%. The value of $\sigma$ approaches zero, implying a negligible level of uncertainties in the time dimension of SAPF degradation. The measurement error $\sigma_\varepsilon$ is 0.3609, which exceeds the instrument maximum testing deviation of 0.3. This is probably attributed to the unavoidable manual positioning of the SAPF during SAR measurement, leading to human-induced measurement errors. As shown in Fig. 5, 95% of the $\alpha_3$ estimates are below zero, which further supports the credibility of the result that high temperatures suppress the degradation of SAPFs. Additionally, 66% of the $T_{threshold}$ estimates fall within the 55-65 ℃ range, with a clear peak at 64 ℃, reflecting the robustness of the temperature threshold findings.

Besides, to validate the robustness of the TERIME-based statistical analysis method, the parameters are independently estimated 20 times based on all degradation data. Then, the mean value, standard deviation and coefficient of variation of the estimates are calculated, as listed in Table III. It can be seen that in addition to $\sigma_a$ and $\sigma$, other parameter estimations are almost consistent in each run, proving that the TERIME-based statistical analysis method ensures robust parameter estimation for the proposed model. Since the influences of $\sigma_a$ and $\sigma$ is minimal, the differences in their values are reasonable and acceptable.

Additionally, the estimation results obtained via the Nelder-Mead method across 20 independent runs are presented in Table IV, utilizing MATLAB's "fminsearchbnd" function [52]. The initial values for parameter estimation are randomly selected within the defined parameter bounds, which are identical to those used in the TERIME-based method. It is evident that, compared to the TERIME-based statistical analysis approach, the estimates obtained using the Nelder-Mead method exhibit significant instability. This instability stems from the large number of unknown parameters in the proposed degradation model and the inherent complexity of the optimization process.

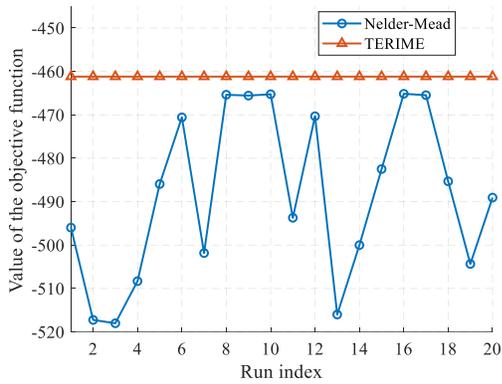

Fig. 7 Variation of the objective function value for two parameter estimation methods across 20 independent runs.

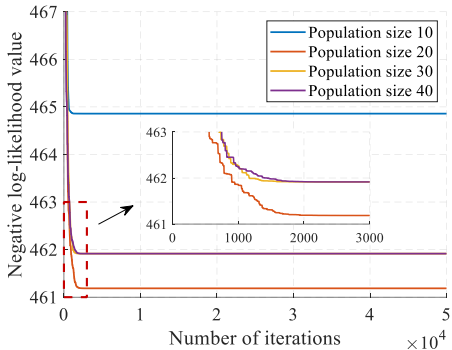

Fig. 8 Variation of negative log-likelihood values under different combinations of population sizes and iteration numbers for the TERIME-based method.

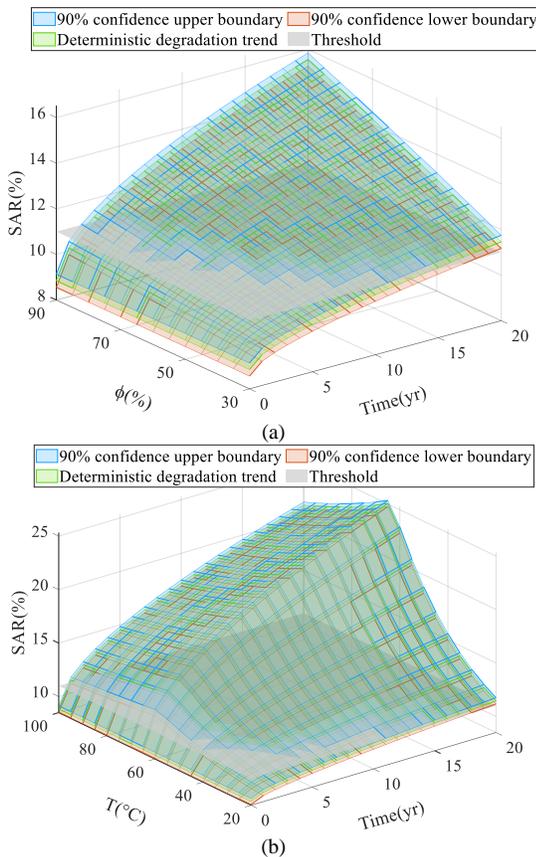

Fig. 9 Deterministic degradation trends and boundaries of SAR: (a) with $\phi$ and time at $T = 20$ ℃ (b) with $T$ and time at $\phi = 30\%$.

The Nelder-Mead method is highly sensitive to initial values and prone to getting trapped in local optima. In contrast, TERIME avoids this issue through enhanced exploration and exploitation strategies, thereby ensuring the accuracy and robustness of the estimation results.

Furthermore, the variation of the objective function value for the two parameter estimation methods across 20 independent runs is illustrated in Fig. 7. Since the objective function used here is the log-likelihood function given by (14), a larger value indicates a better parameter estimation performance. It can be observed that the TERIME-based method yields nearly identical and consistently high likelihood values over all runs. In contrast, the likelihood values obtained using the Nelder-Mead method vary significantly with the run index. Therefore, the TERIME-based statistical analysis method achieves more accurate and more reliable parameter estimation results compared to the Nelder–Mead method.

Fig. 8 presents the variation of negative log-likelihood values under different combinations of population sizes and iteration numbers for the TERIME-based method. The results show that regardless of the population size, convergence typically occurs near 2000 iterations. Thus, setting the number of iterations to 50000 ensures sufficient convergence. Additionally, the algorithm performs best with a population size of 20, which is consistent with the sensitivity analysis result reported in [47]. These findings support the rationality of the chosen hyper-parameters for accurate parameter estimation.

Based on the parameter estimation results in Table II, the degradation predictions under all the stress levels can be obtained according to Algorithm 2, which are shown in Fig. 6. The results demonstrate that the proposed degradation model effectively describes the degradation trends across various stress levels, with the 90% confidence intervals covering most of the measurement data. Notably, the degradation intervals presented in Fig. 6 stem from inherent uncertainty of the degradation process for a batch of SAPFs and measurement errors, and are non-relevant to the parameter estimation uncertainty presented in Tables II and III, which originates from insufficient sample size or the estimation technique.

Fig. 9 illustrates the performance degradation of SAPF with varying $T$, $\phi$, and $t$. As $\phi$ increases, the performance degradation of SAPF accelerates, which in turn reduces its reliability. On the other hand, when $T$ rises, the performance degradation of SAPF first accelerates and then decelerates due to the degradation mechanism transition.

### IV. COMPARATIVE STUDIES AND RELIABILITY PREDICTION

#### A. Comparative Studies

In order to verify the effectiveness of the proposed stress model in describing the stress-induced degradation pattern of SAPFs, and determine *the* most suitable degradation models with uncertainties, comparative studies are conducted. The proposed degradation model expressed by (10) is designated as $M_0$. The following degradation models are selected for comparison:

1) The widely used degradation model which assumes unchanged degradation mechanism with stress is compared [31-36], incorporating the time dimension uncertainty, unit-to-unit variability (i.e., uncertainty in the degradation rate, also known as random effect), initial performance uncertainty and measurement error. Notably, this degradation model is also termed as "accelerated degradation model" in the literature, which is expressed by (15) and is designated as $M_1$.

2) As shown in Table II, the estimates associated with the time dimension uncertainty and unit-to-unit variability are relatively small, whereas measurement error has a more significant impact. Consequently, two ablated versions of the proposed degradation model are considered for comparison: one that omits both time dimension uncertainty and unit-to-unit variability (designated as $M_2$), and another that omits measurement error (designated as $M_3$). Specifically, $M_2$ is expressed by:

$$Y(t,s) = Y_0 + ae(s)t^\beta + \varepsilon,$$

$$e(s) = \begin{cases} \exp(\alpha_1 \varphi_1(T) + \alpha_2 \phi^*), & T < T_{threshold} \\ \exp(\alpha_1 + \alpha_2 \phi^* + \alpha_3 \varphi_3(T)), & T \geq T_{threshold} \end{cases}, \quad (18)$$

$$Y_0 \sim N(\mu_{Y_0}, \sigma_{Y_0}^2), \varepsilon \sim N(0, \sigma_\varepsilon^2).$$

$M_3$ is expressed by:

$$Y(t,s) = Y_0 + ae(s)t^\beta + \sigma B(t),$$

$$e(s) = \begin{cases} \exp(\alpha_1 \varphi_1(T) + \alpha_2 \phi^*), & T < T_{threshold} \\ \exp(\alpha_1 + \alpha_2 \phi^* + \alpha_3 \varphi_3(T)), & T \geq T_{threshold} \end{cases}, \quad (19)$$

$$Y_0 \sim N(\mu_{Y_0}, \sigma_{Y_0}^2), a \sim N(\mu_a, \sigma_a^2).$$

By comparing $M_0$ with $M_1$, the superiority of the proposed stress model can be demonstrated. Furthermore, by comparing $M_0$ with $M_2$ and $M_3$, the most suitable degradation models with uncertainties for SAPFs can be determined.

*1) Model Fitting Comparison*: Initially, the comparison between the two models focuses on their ability to fit degradation data across all stress levels, including both deterministic degradation trends and uncertainties. Reliability predictions can only be considered credible if the model predicts the degradation trends across varying stress levels accurately and quantifies uncertainties effectively.

For deterministic degradation trends, the root mean square error (RMSE) is selected as metrics to evaluate the alignment between the predicted and real degradation trends, which is expressed as:

$$\text{RMSE} = \sqrt{\sum_{l=1}^{k}\sum_{j=1}^{m_{li}} \left(\bar{y}_{lj} - \hat{\bar{y}}_{lj}\right)^2 / \sum_{l=1}^{k} m_l}, i=1,2,...,n_l, \quad (20)$$

where $\bar{y}_{lj}$ is the $j^{th}$ real average degradation value at the $l^{th}$ stress level calculated by:

TABLE V
PARAMETER ESTIMATION RESULTS OF DIFFERENT MODELS

| Parameters | $M_0$ | $M_1$ | $M_2$ | $M_3$ |
|---|---|---|---|---|
| $\mu_{Y0}$ | 8.844 | 8.836 | 8.844 | 8.845 |
| $\sigma_{Y0}$ | 0.1754 | 0.1705 | 0.1754 | 0.3361 |
| $\mu_a$ | 0.003374 | 0.01063 | 0.003374 | 0.0004123 |
| $\sigma_a$ | 1.590e-11 | 2.903e-10 | / | 9.912e-11 |
| $\alpha_1$ | 1.836 | 0.776 | 1.836 | 4.865 |
| $\alpha_2$ | 0.9439 | 1.228 | 0.9439 | 6.475e-15 |
| $\alpha_3$ | -0.2901 | / | -0.2901 | -1.017 |
| $\beta$ | 0.5510 | 0.4618 | 0.5510 | 0.5873 |
| $\sigma$ | 4.081e-10 | 1.189e-9 | / | 0.1483 |
| $\sigma_\varepsilon$ | 0.3609 | 0.3628 | 0.3609 | / |
| $T_{threshold}$ | 63.75 | / | 63.75 | 65.00 |

TABLE VI
COMPARISON RESULTS OF MODEL FITTING

| Models | RMSE | $L_{max}$ | AIC |
|---|---|---|---|
| $M_0$ | **0.217** | **-461.19** | 944.38 |
| $M_1$ | 0.219 | -464.86 | 947.72 |
| $M_2$ | **0.217** | **-461.19** | **940.38** |
| $M_3$ | 0.245 | -719.41 | 1458.82 |

TABLE VII
STRESS LEVELS ASSIGNED FOR PREDICTION IN EACH TEST

| Test number | $T$ | $\phi$ | Test number | $T$ | $\phi$ |
|---|---|---|---|---|---|
| 1 | 40 °C | 50% | 5 | 60 °C | 90% |
| 2 | 40 °C | 75% | 6 | 65 °C | 75% |
| 3 | 50 °C | 50% | 7 | 85 °C | 80% |
| 4 | 50 °C | 75% | 8 | 90 °C | 85% |

$$\bar{y}_{lj} = \sum_{i=1}^{n_l} y_{lij} / n_l, \quad (21)$$

and $\hat{\bar{y}}_{lj}$ is the $j^{th}$ predicted average degradation value at the $l^{th}$ stress level derived by the degradation model.

Regarding uncertainties, the maximum value of the log-likelihood function $L_{max}$ and the Akaike information criterion (AIC) are chosen as measures to describe the fitting effectiveness of the model for all data. $L_{max}$ is calculated by (14) using the estimated model parameters. AIC is calculated by:

$$\text{AIC} = -2L_{max} + 2n_p, \quad (22)$$

where $n_p$ is the number of unknown parameters.

The unknown parameters in these degradation models can be estimated according to Algorithm 1, which are listed in Table V. Then, the fitting performance results for the degradation models are presented in Table VI. The proposed model $M_0$ slightly outperforms the existing model $M_1$ in both deterministic degradation prediction and uncertainty quantification, with RMSE reduced by 0.9%, $L_{max}$ improved by 0.8%, and AIC decreased by 3.5%, highlighting the effectiveness of the proposed stress model that accounts for mechanism transitions. Model $M_2$ yields the same RMSE and $L_{max}$ values as $M_0$, suggesting that uncertainties in the time and unit dimensions are negligible in the degradation of SAPFs, which aligns with the parameter estimation results of $M_0$. In contrast, the fitting performance of $M_3$ is significantly poorer, with RMSE increased by 12.9% and AIC increased by 54.5%, indicating that measurement error cannot be ignored in the degradation modeling of SAPFs.

*2) Model Extrapolation Comparison*: Notably, one objective of this study is to utilize the available degradation data to predict reliability at other stress levels. Consequently, evaluating the model's extrapolation performance at unseen

TABLE VIII
COMPARISON RESULTS OF MODEL EXTRAPOLATION

| Test number | $M_0$ | | $M_1$ | | $M_2$ | | $M_3$ | |
|---|---|---|---|---|---|---|---|---|
| | RMSE | $L_{max}$ | RMSE | $L_{max}$ | RMSE | $L_{max}$ | RMSE | $L_{max}$ |
| 1 | 0.145 | **-39.9** | **0.140** | -40.4 | 0.145 | **-39.9** | 0.180 | -75.6 |
| 2 | 0.221 | **-68.2** | **0.216** | -69.1 | 0.221 | **-68.2** | 0.242 | -101.1 |
| 3 | 0.333 | **-107.2** | **0.327** | -107.9 | 0.333 | **-107.2** | **0.327** | -179.3 |
| 4 | **0.258** | **-73.5** | 0.259 | -73.6 | **0.258** | **-73.5** | 0.289 | -125.5 |
| 5 | 0.150 | **-33.7** | **0.142** | -34.8 | 0.150 | **-33.7** | 0.225 | -40.9 |
| 6 | **0.098** | **-21.1** | 0.165 | -22.9 | **0.098** | **-21.1** | 0.391 | -51.3 |
| 7 | **0.167** | **-71.7** | 0.172 | -72.3 | **0.167** | **-71.7** | 0.176 | -94.0 |
| 8 | 0.240 | -57.7 | **0.235** | **-57.6** | 0.240 | -57.7 | 0.340 | -68.3 |
| Mean | **0.202** | **-59.1** | 0.207 | -59.8 | **0.202** | **-59.1** | 0.271 | -92.0 |

*Note*: The best results for different metrics of each test are highlighted in bold.

TABLE IX
COMPARISON RESULTS OF MODEL ROBUSTNESS

| Test number | $M_0$ | | $M_1$ | | $M_2$ | | $M_3$ | |
|---|---|---|---|---|---|---|---|---|
| | KLD | CMD | KLD | CMD | KLD | CMD | KLD | CMD |
| 1 | **0.92** | **3.74** | 7.77 | 14.83 | **0.92** | **3.74** | 16.62 | 29.52 |
| 2 | **0.96** | **3.41** | 11.50 | 17.66 | **0.96** | **3.41** | 23.84 | 34.77 |
| 3 | **2.45** | **14.42** | 14.43 | 72.36 | **2.45** | **14.42** | 30.03 | 137.58 |
| 4 | **0.57** | **4.36** | 1.25 | 9.81 | **0.57** | **4.36** | 2.31 | 15.26 |
| 5 | **1.90** | **6.69** | 13.72 | 21.14 | **1.90** | **6.69** | 28.73 | 36.89 |
| 6 | **0.26** | **1.57** | 0.78 | 2.80 | **0.26** | **1.57** | 1.52 | 4.36 |
| 7 | **0.07** | **0.41** | 0.36 | 2.43 | **0.07** | **0.41** | 0.88 | 4.64 |
| 8 | **6.97** | **59.33** | 13.23 | 100.50 | **6.97** | **59.33** | 26.77 | 155.21 |
| Mean | **1.76** | **11.74** | 7.88 | 30.19 | **1.76** | **11.74** | 16.34 | 52.28 |

*Note*: The best results for different metrics of each test are highlighted in bold.

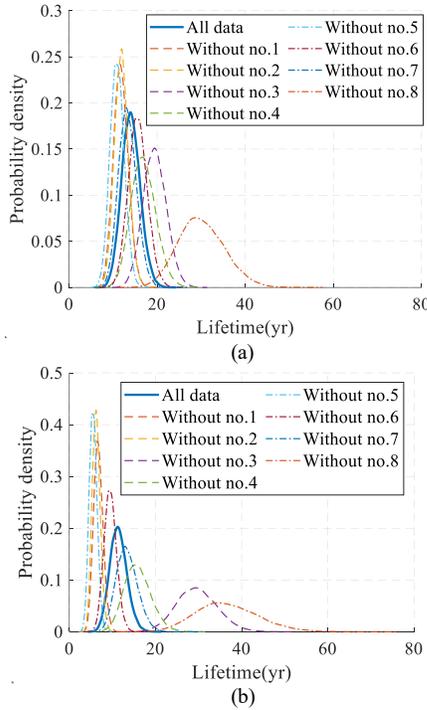

Fig. 10 Comparison of the lifetime distribution using all degradation data with those using the degradation data without the specific test stress level: (a) the proposed model (model $M_0$); (b) the traditional model with unchanged degradation mechanism (model $M_1$).

stress levels is essential. For this purpose, model extrapolation performance verification tests are conducted. Since the degradation experiment involves 8 stress levels, 8 validation tests are set. In each test, the degradation data under one specific stress level are used for prediction, whereas the data from other stress levels are used to establish the model, thereby evaluating the model's extrapolation capability. The stress levels assigned for prediction in each test are outlined in Table VII.

Table VIII presents the degradation prediction results of the degradation models for each test. It can be seen that across all the tests, model $M_1$ is sometimes superior in deterministic degradation predictions, whereas model $M_0$ is always superior in describing all the degradation data considering uncertainties. On average, $M_0$ achieves a 2.4% lower RMSE and a 1.2% higher $L_{max}$ compared to $M_1$, indicating superior extrapolation capability. Model $M_2$ yields identical RMSE and $L_{max}$ values to $M_0$, reaffirming that uncertainties in time and unit dimensions have negligible impact. $M_3$ performs significantly worse, with RMSE increased by 34.2% and $L_{max}$ decreased by 35.9%, highlighting the necessity of accounting for measurement error in SAPF degradation modeling.

*3) Model Robustness Comparison*: In addition to model fitting and extrapolation performance, robustness is another key aspect of model credibility. If the degradation model has accurately captured the performance degradation law based on existing data, introducing degradation data from new stress levels should cause only minor variations in the reliability predictions. Conversely, if the model fails to describe the degradation law effectively, introducing degradation data from new stress levels would cause significant prediction changes.

To evaluate the robustness of the model, the test settings in Table VI are employed again. The variation in predictions at the specific test stress level are assessed by comparing the result of the degradation model fitted by all the degradation data with that fitted by the data without the specific test stress level. The Kullback-Leibler divergence (KLD) of the lifetime distributions and the Cramér-von Mises distance (CMD) of the reliability distributions are selected as metrics, which are expressed by (23) and (24), respectively.

$$\text{KLD} = \sum_{i=1}^{K} f_{all}(t_i) \ln \frac{f_{all}(t_i)}{f_{test}(t_i)}, \quad (23)$$

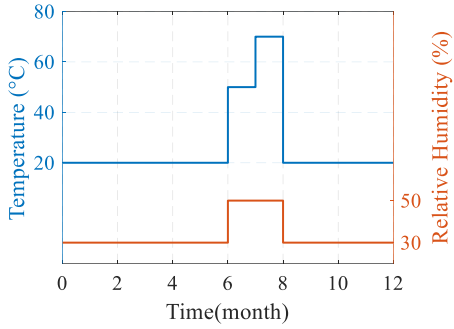

Fig. 11  The storage condition of the SAPF in one year.

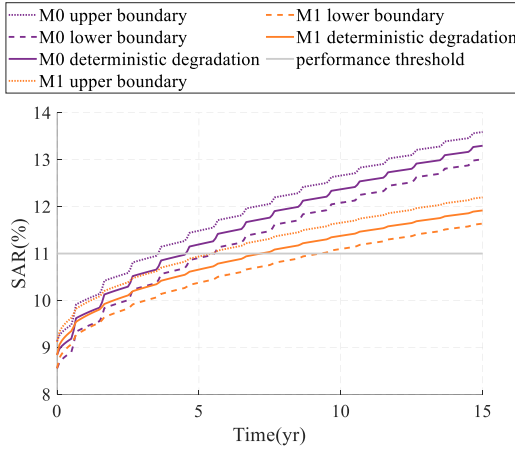

Fig. 12  Performance degradation predictions of SAPF obtained by both models.

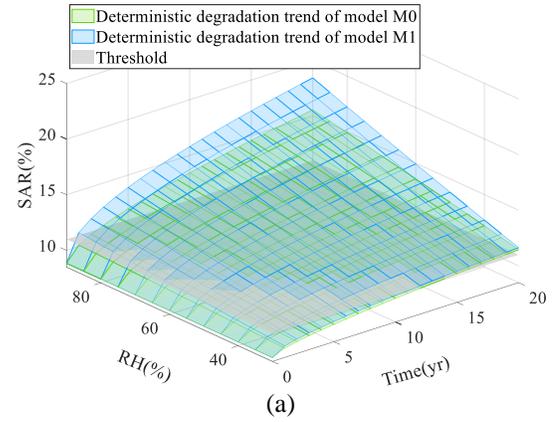

(a)

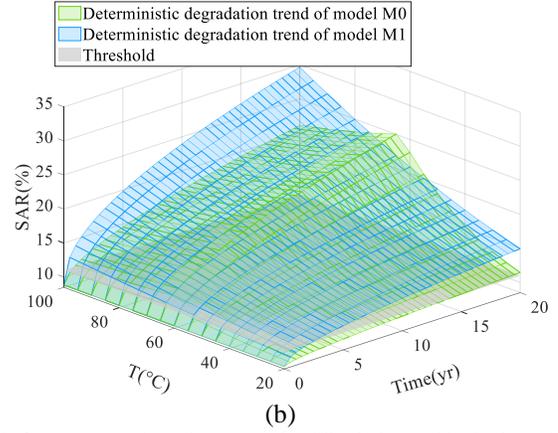

(b)

Fig. 13  Deterministic degradation trends of SAR obtained by both models: (a) with $\phi$ and time at $T = 20$ ℃ (b) with $T$ and time at $\phi = 30\%$.

$$\text{CMD} = \sum_{i=1}^{K} \left( R_{all}(t_i) - R_{test}(t_i) \right)^2, \quad (24)$$

where $t_i$ represents the $i^{th}$ discrete time points; $K$ is the total number of time points; $f_{all}$ and $R_{all}$ are the predicted lifetime and reliability distributions of the model fitted by all the degradation data; $f_{test}$ and $R_{test}$ are the predicted lifetime and reliability distributions of the model fitted by the degradation data without the specific test stress level. In this study, $K$ is set as 500. The reliability predictions are conducted at a specific stress level $T = 20$ ℃ and $\phi = 30\%$.

Table IX gives the robustness evaluation results, with Fig. 10 illustrating the lifetime distribution comparisons of $M_0$ and $M_1$. It can be seen that the model $M_0$ is more robust than the model $M_1$, mainly because the traditional degradation model fails to capture the real degradation pattern of temperature. On average, $M_0$ reduces KLD by 77.6% and CMD by 61.1% compared to $M_1$, indicating much better consistency under varying conditions. Again, model $M_2$ shows identical performance to $M_0$, while $M_3$ performs the worst, with KLD and CMD increased by 828% and 345% respectively, confirming the necessity of accounting for measurement error.

### B. Reliability Prediction

In practical applications, the storage condition of SAPFs varies in response to the product demands. According to the requirements of an aerospace system, the storage condition of the SAPF over the span of a year is depicted in Fig. 11.

Based on the given storage condition given in Fig. 11, the performance degradation of SAPFs can be predicted by the models $M_0$ and $M_1$ as shown in Fig. 12. The results show that the failure predicted by $M_0$ occurs earlier than that predicted by $M_1$. To provide a comprehensive understanding of this result, the impacts of $\phi$ and $T$ on deterministic degradation trend predictions derived by both models are displayed in Fig. 13.

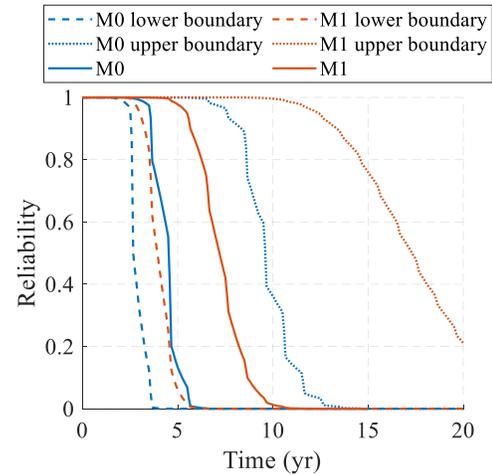

Fig. 14  Reliability predictions of SAPF obtained by both models with intervals at 90% confidence level.

From Fig. 13, when $T = 20$ ℃ and $\phi = 30\%$, the degradation predicted by $M_0$ is slower than that by $M_1$, leading to a slower degradation prediction for $M_0$ in Fig. 12 initially. However, when $T = 50$ ℃ or $T = 70$ ℃, it indicates that the degradation predicted by $M_1$ is significantly slower than that of $M_0$. Therefore, the degradation predicted by $M_0$ in Fig. 12 eventually surpasses that of $M_1$.

TABLE X
STORAGE CONDITIONS CONSIDERED FOR SAPF IN THE WAREHOUSE

| Storage condition number | $T$ | $\phi$ |
|---|---|---|
| 1 | 20 °C | 30% |
| 2 | 20 °C | 40% |
| 3 | 25 °C | 30% |
| 4 | 25 °C | 40% |

TABLE XI
STORAGE CONDITIONS CONSIDERED FOR SAPF DURING LOGISTICS

| Logistic condition number | $T$ | $\phi$ |
|---|---|---|
| 1 | 50 °C, 70 °C | 50% |
| 2 | 40 °C, 60 °C | 60% |

Then, the reliability predictions derived by both models with intervals at 90% confidence level are illustrated in Fig. 14. The results indicate that ignoring the stress-induced degradation mechanism transition will result in an overestimation of SAPF storage reliability. For the proposed degradation model $M_0$, the reliability of SAPF after 5 years of storage based on the point estimates is approaching zero, whereas the compared model $M_1$ presents a reliability above 0.9, potentially causing wrong operation decisions and maintenance strategies, ultimately raising the failure risk of the aerospace system.

In addition, the reliability prediction interval of model $M_1$ is significantly wider than that of model $M_0$. The upper reliability boundary predicted by model $M_0$ approaches zero at around 13 years, whereas that of model $M_1$ remains above 0.2 even after 20 years. A wider prediction interval indicates greater uncertainty, which can amplify decision-making risks. This increased uncertainty is attributed to the inadequacy of model $M_1$ in capturing the stress-induced degradation mechanism transition of SAPFs. Besides, the lower reliability boundary for model $M_0$ reaches about 0.99 at 2 years and declines to nearly zero by 4 years. This information can provide guidance for decision-making in extreme scenarios.

## V. DISCUSSIONS

### A. Relationship between Storage Degradation and Operation in Space

The degradation mechanisms investigated in this study correspond to the ground-storage stage of SAPFs, during which temperature and humidity induce slow chemical corrosion of the aluminum layer, leading to a gradual increase in SAR. Previous studies have also shown that SAPFs exposed to the space environment undergo performance degradation [5-8]. However, the dominant mechanism in space is radiation-induced chemical modification of the polyimide substrate rather than corrosion of the aluminum layer. Processes such as molecular bond scission, formation of color centers and optical darkening have been reported to increase SAR.

Despite the distinct physical origins of degradation during ground storage and space exposure, their consequence is the same, as both result in increased SAR. Therefore, degradation accumulated during storage directly affects the initial state of SAPF before launch. If SAR has already increased during storage, the SAPF enters the space environment with a reduced thermal control margin, thereby shortening its operational lifetime. In this context, the storage degradation model developed in this study establishes a baseline condition required for accurate reliability and lifetime assessments in space.

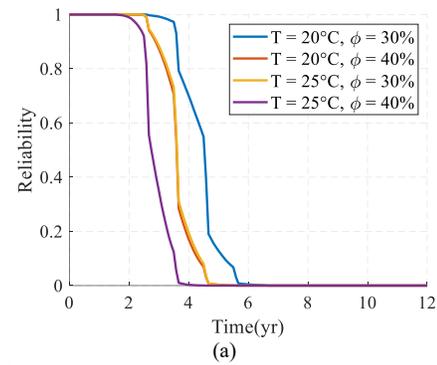

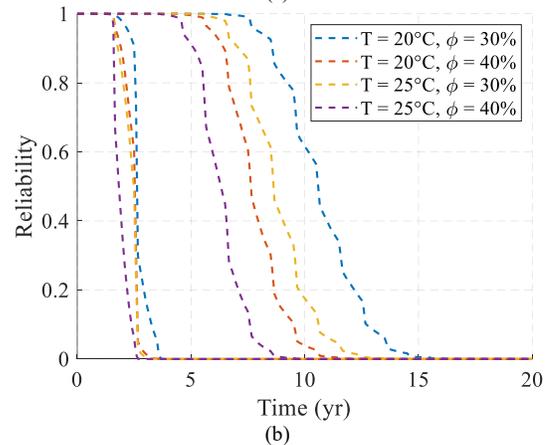

Fig. 15 Reliability predictions of SAPF under different storage conditions: (a) point predictions; (b) interval predictions at 90% confidence level.

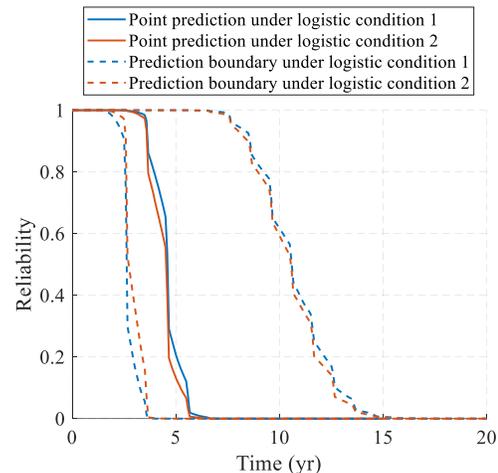

Fig. 16 Reliability predictions of SAPF under different logistic conditions with intervals at 90% confidence level.

### B. Guidance for Storage Condition Determination

According to the storage profiles given in Fig. 11, SAPF is stored outside the warehouse in the seventh and eighth months to meet specific logistic demands, and it remains in a controlled warehouse environment during the rest of the year. Although the storage conditions in the warehouse can minimize degradation by controlling temperature and humidity, this also leads to increased costs. A critical issue is how to define appropriate storage conditions based on reliability requirements. Herein, four distinct warehouse storage conditions are considered as presented in Table X. Then, the reliability predictions under different storage conditions with intervals at 90% confidence level are illustrated in Fig. 15. For a high reliability requirement of 0.99, the storage duration is 3 years under Condition 1, 2.5 years under Conditions 2 and 3, and 2 years under Condition 4. For the lower reliability boundary in extreme scenarios, Condition 1 allows for 1.75 years of storage, while the other conditions support 1.5 years. These results can provide practical insights for establishing appropriate storage conditions in the warehouse.

In addition, to reflect the impact of storage conditions during different logistics, two logistic conditions are considered as listed in Table XI. Fig. 16 illustrates the reliability predictions under these conditions with 90% confidence intervals. It is evident that both logistic conditions yield similar reliability results. Logistics 1 shows slightly faster performance degradation compared to Logistics 2 but lower uncertainty. Overall, this analysis demonstrates that the proposed model can predict reliability under varying storage and logistic conditions, thereby informing the selection of suitable storage profiles.

## VI. CONCLUSION

SAPFs play a key role in thermal management of aerospace systems. In existing studies, the performance degradation of SAPF in ground environments during storage is neglected. To address this issue, in this work, a parameter analysis method is presented to determine the degradation mechanism of SAPFs based on experimental data, and a novel degradation model is proposed to describe the stress-induced degradation mechanism transition. The results and comparative studies demonstrate that:

1) The proposed parameter analysis method successfully discovers two distinct stages in the impact of temperature on SAR degradation from the experimental data, demonstrating that SAPFs follow similar degradation mechanism transition observed in bulk aluminum.
2) The temperature threshold for the degradation mechanism transition estimated by the proposed model falls within the range of 54.30 ℃ to 75.76 ℃ at 90% confidence level, aligning closely with the experimental results for aluminum corrosion reported in the literature [13] (approximately 60 ℃).
3) According to multiple runs of parameter estimation, the standard deviations of all the estimates are below 1e-6, proving that the proposed TERIME-based statistical analysis method can give robust parameter estimates.
4) For the performance degradation of SAPFs, unit-to-unit variability and the uncertainty in the time dimension are negligible, while the uncertainty in the initial performance is significant. Therefore, reducing the uncertainty in the initial performance is the primary design measure for improving its reliability.

In the future, the proposed degradation model could be extended to other products that exhibit stress-induced degradation mechanism transitions. Additionally, investigating such transitions within the framework of non-Arrhenius models [53] represents a promising research direction. As for applications in SAPFs, the interaction effects of temperature and humidity can be explored in the future for further enhancing model accuracy [37]. Moreover, the geometries and manufacturing conditions of SAPFs will be incorporated in the degradation model to guide the reliability design of SAPFs.

## REFERENCES

[1] H. Huang, and C. Bu, "Design and Verification of Thermal Control System of Communication Satellite," *Aerospace*, vol. 11, no. 10, pp. 803. 2024.
[2] M. Hołyńska, A. Tighe, and C. Semprimoschnig, "Coatings and thin films for spacecraft thermo-optical and related functional applications," *Adv. Mater. Interfaces.*, vol. 5, no. 11, pp. 1701644. 2018.
[3] R. You, W. Gao, C. Wu, H. Li, R. You, W. Gao, C. Wu, and H. Li, "Reliability Design of Spacecraft Antenna," *Technologies for Spacecraft Antenna Engineering Design*, pp. 249-292. 2021.
[4] A. K. Sharma, and N. Sridhara, "Degradation of thermal control materials under a simulated radiative space environment," *Adv. Space Res.*, vol. 50, no. 10, pp. 1411-1424. 2012.
[5] C. Li, D. Yang, S. He, and M. Mikhailov, "Effect of electron exposure on optical properties of aluminized polyimide film," *J. Mater. Res.*, vol. 17, no. 9, pp. 2442-2446. 2002.
[6] V. Milinchuk, E. Klinshpont, O. Anan'eva, and O. Pasevich, "Characterization of polyimide films exposed at the mir orbital space station," *High Energy Chem.*, vol. 48, pp. 1-4. 2014.
[7] J. A. Dever, R. Messer, C. Powers, J. Townsend, and E. Wooldridge, "Effects of vacuum ultraviolet radiation on thin polyimide films," *High Perform. Polym.*, vol. 13, no. 3, pp. S391-S400. 2001.
[8] M. Dembska, T. Renger, and M. Sznajder, "Thermo-Optical Property Degradation of ITO-Coated Aluminized Polyimide Thin Films Under VUV and Low-Energy Proton Radiation," *Metall. Mater. Trans. A.*, vol. 51, no. 9, pp. 4922-4929. 2020.
[9] X.-S. Si, C.-H. Hu, X. Kong, and D.-H. Zhou, "A residual storage life prediction approach for systems with operation state switches," *IEEE Trans. Ind. Electron.*, vol. 61, no. 11, pp. 6304-6315. 2014.
[10] G. Garnier, Y. Brechet, and L. Flandin, "Development of an experimental technique to assess the permeability of metal coated polymer films," *J. Mater. Sci.*, vol. 44, pp. 4692-4699. 2009.
[11] C. Zhou, W. Li, G. Li, and Y. Niu, "Hydrothermal Aging Mechanism of Aluminized Polyethylene Terephthalate Films," *Polym. Mater. Sci. Eng.*, vol. 33, no. 3, pp. 64-70. 2017.
[12] C. Leygraf, I. O. Wallinder, J. Tidblad, and T. Graedel, "Appendix C: The Atmospheric Corrosion Chemistry of Aluminum," *Atmospheric Corrosion*, pp. 272-281, 2016.
[13] M. Cao, L. Liu, L. Fan, Z. Yu, Y. Li, E. E. Oguzie, and F. Wang, "Influence of Temperature on Corrosion Behavior of 2A02 Al Alloy in Marine Atmospheric Environments," *Materials*, vol. 11, no. 2, pp. 235. 2018.
[14] Y. Xie, A. Leong, J. Zhang, and J. J. Leavitt, "Aluminum alloy corrosion in boron‐containing alkaline solutions," *Mater. Corros.*, vol. 70, no. 5, pp. 810-819. 2019.
[15] S. L'Haridon-Quaireau, M. Laot, K. Colas, B. Kapusta, S. Delpech, and D. Gosset, "Effects of temperature and pH on uniform and pitting corrosion of aluminium alloy 6061-T6 and characterisation of the hydroxide layers," *J. Alloys Compd.*, vol. 833, pp. 155146. 2020.
[16] Q. Li, Y. Zhang, Y. Cheng, X. Zuo, Y. Wang, X. Yuan, and H. Huang, "Effect of temperature on the corrosion behavior and corrosion resistance of copper-aluminum laminated composite plate," Materials, vol. 15, no. 4, pp. 1621. 2022.
[17] X. Li, B. Zhang, R. Ren, L. Li, and S. P. Simonovic, "Spatio-temporal heterogeneity of climate warming in the Chinese Tianshan Mountainous Region," Water, vol. 14, no. 2, pp. 199. 2022.
[18] M. Saif, S. Zhang, A. Haque, and K. J. Hsia, "Effect of native Al2O3 on the elastic response of nanoscale Al films," *Acta Mater.,* vol. 50, no. 11, pp. 2779-2786. 2002.
[19] G. F. Harrington, and J. Santiso, "Back-to-Basics tutorial: X-ray diffraction of thin films," *J. Electroceram.,* vol. 47, no. 4, pp. 141-163. 2021.
[20] A. Abou Chaaya, R. Viter, M. Bechelany, Z. Alute, D. Erts, A. Zalesskaya, K. Kovalevskis, V. Rouessac, V. Smyntyna, and P. Miele, "Evolution of microstructure and related optical properties of ZnO grown by atomic layer deposition," *Beilstein J. Nanotechnol.,* vol. 4, no. 1, pp. 690-698. 2013.
[21] X. Xi, M. Chen, H. Zhang, and D. Zhou, "An improved non-Markovian degradation model with long-term dependency and item-to-item uncertainty," *Mech. Syst. Signal Process.*, vol. 105, pp. 467-480. 2018.
[22] A. Asgari, W. Si, W. Wei, K. Krishnan, and K. Liu, "Multivariate degradation modeling using generalized cauchy process and application in life prediction of dye-sensitized solar cells," *Reliab. Eng. Syst. Saf.*, vol. 255, pp. 110651. 2025.

[23] A. Asgari, W. Si, L. Yuan, K. Krishnan, and W. Wei, "Multivariable degradation modeling and life prediction using multivariate fractional Brownian motion," *Reliab. Eng. Syst. Saf.*, vol. 248, pp. 110146. 2024.

[24] X. Xi, M. Chen, and D. Zhou, "Remaining useful life prediction for degradation processes with memory effects," *IEEE Trans. Reliab.*, vol. 66, no. 3, pp. 751-760. 2017.

[25] K. Wang, T. Zhao, Y. Yuan, Z. Hao, Z. Chen, and H. Dui, "A new multi-layer performance analysis of unmanned system-of-systems within IoT," *Reliab. Eng. Syst. Saf.*, vol. 259, pp. 110953. 2025.

[26] S. Gao, F. Sun, X. Zhao, and Y. Li, "Optimal warranty period design for new products subject to degradation and environmental shocks considering imperfect maintenance," *Reliab. Eng. Syst. Saf.*, vol. 256, pp. 110710. 2025.

[27] J. Zhou, Z. Li, H. Nassif, and D. W. Coit, "A two-stage Weibull-gamma degradation model with distinct failure mechanism initiation and propagation stages," *Reliab. Eng. Syst. Saf.*, vol. 256, pp. 110773. 2025.

[28] H. Zheng, J. Yang, and Y. Zhao, "Reliability demonstration test plan for degraded products subject to Gamma process with unit heterogeneity," *Reliab. Eng. Syst. Saf.*, vol. 240, pp. 109617. 2023.

[29] Z. Z. Yan, L. J. Sun, and Y. Y. Liang, "Remaining Useful Life Analysis for Two-Stage Nonlinear Inverse Gaussian Process With Random Effects," *Qual. Reliab. Eng. Int.*, vol. 41, no. 4, pp. 1447-1460. 2025.

[30] T.-H. Fan, Y.-S. Dong, and C.-Y. Peng, "A complete Bayesian degradation analysis based on inverse Gaussian processes," *IEEE Trans. Reliab.*, vol. 73, no. 1, pp. 536-548. 2023.

[31] J. Cai, W. Q. Meeker, and Z.-S. Ye, "Optimal Planning of Destructive Degradation Tests," *Technometrics*, pp. 1-12. 2025.

[32] S.-S. Chen, X.-Y. Li, and W.-R. Xie, "Reliability modeling and statistical analysis of accelerated degradation process with memory effects and unit-to-unit variability," *Appl. Math. Model.*, vol. 138, pp. 115788. 2025.

[33] W.-B. Chen, X.-Y. Li, and R. Kang, "Attractor based performance characterization and reliability evolution for electromechanical systems," *Mech. Syst. Signal Process.*, vol. 222, pp. 111803. 2025.

[34] Y. Li, H. Gao, H. Chen, C. Liu, Z. Yang, and E. Zio, "Accelerated degradation testing for lifetime analysis considering random effects and the influence of stress and measurement errors," *Reliab. Eng. Syst. Saf.*, vol. 247, pp. 110101. 2024.

[35] X.-Y. Li, J.-P. Wu, L. Liu, M.-L. Wen, and R. Kang, "Modeling accelerated degradation data based on the uncertain process," *IEEE Trans. Fuzzy Syst.*, vol. 27, no. 8, pp. 1532-1542. 2018.

[36] W. Si, Y. Shao, and W. Wei, "Accelerated degradation testing with long-term memory effects," *IEEE Trans. Reliab.*, vol. 69, no. 4, pp. 1254-1266. 2020.

[37] X. Ye, Y. Hu, B. Zheng, C. Chen, and G. Zhai, "A new class of multi-stress acceleration models with interaction effects and its extension to accelerated degradation modelling," *Reliab. Eng. Syst. Saf.*, vol. 228, pp. 108815. 2022.

[38] X. Cao, and K. Peng, "Multi-phase Degradation Modeling and Remaining Useful Life Prediction Considering Aleatory and Epistemic Uncertainty," *IEEE Sens. J.*, vol. 23, no. 22, pp. 27757-27770. 2023.

[39] H. Zheng, J. Yang, and Y. Zhao, "Reliability analysis of multi-stage degradation with stage-varying noises based on the nonlinear Wiener process," *Appl. Math. Model.*, vol. 125, pp. 445-467. 2024.

[40] Z. Wang, P. Jiang, Z. Chen, Y. Li, W. Ren, L. Dong, W. Du, J. Wang, X. Zhang, and H. Shi, "Remaining useful life prediction method based on two-phase adaptive drift Wiener process," *Reliab. Eng. Syst. Saf.*, pp. 110908. 2025.

[41] X. Cai, N. Li, and M. Xie, "RUL prediction for two-phase degrading systems considering physical damage observations," *Reliab. Eng. Syst. Saf.*, vol. 244, pp. 109926. 2024.

[42] H. Zhang, X. Xi, and R. Pan, "A two-stage data-driven approach to remaining useful life prediction via long short-term memory networks," *Reliab. Eng. Syst. Saf.*, vol. 237, pp. 109332. 2023.

[43] X. Liu, X. Wang, M. Xie, and Z. Ye, "Robust degradation state identification in the presence of parameter uncertainty and outliers," *IEEE Trans. Ind. Inform.*, vol. 20, no. 2, pp. 2644-2652. 2023.

[44] Z.-S. Ye, L.-P. Chen, L. C. Tang, and M. Xie, "Accelerated degradation test planning using the inverse Gaussian process," *IEEE Trans. Reliab.*, vol. 63, no. 3, pp. 750-763. 2014.

[45] J. Li, Z. Wang, X. Liu, Y. Zhang, H. Fu, and C. Liu, "A Wiener process model for accelerated degradation analysis considering measurement errors," *Microelectron. Reliab.*, vol. 65, pp. 8-15. 2016.

[46] J.-Z. Kong, D. Wang, T. Yan, J. Zhu, and X. Zhang, "Accelerated stress factors based nonlinear Wiener process model for lithium-ion battery prognostics," *IEEE Trans. Ind. Electron.*, vol. 69, no. 11, pp. 11665-11674. 2021.

[47] S.-S. Chen, Y.-T. Jiang, W.-B. Chen, and X.-Y. Li, "TERIME: An improved RIME algorithm with enhanced exploration and exploitation for robust parameter extraction of photovoltaic models," *J. Bionic Eng.*, vol. 22, pp. 1535-1556. 2025.

[48] R. De Bin, S. Janitza, W. Sauerbrei, and A.-L. Boulesteix, "Subsampling versus bootstrapping in resampling-based model selection for multivariable regression," *Biometrics*, vol. 72, no. 1, pp. 272-280. 2016.

[49] J. A. Hartigan, "Using subsample values as typical values," *J. Am. Stat. Assoc.*, vol. 64, no. 328, pp. 1303-1317. 1969.

[50] S. Janitza, H. Binder, and A. L. Boulesteix, "Pitfalls of hypothesis tests and model selection on bootstrap samples: Causes and consequences in biometrical applications," *Biom. J.*, vol. 58, no. 3, pp. 447-473. 2016.

[51] W. Sauerbrei, A. Buchholz, A. L. Boulesteix, and H. Binder, "On stability issues in deriving multivariable regression models," *Biom. J.*, vol. 57, no. 4, pp. 531-555. 2015.

[52] J. D'Errico, "fminsearchbnd, fminsearchcon," [Online]. Available: hhttps://www.mathworks.com/matlabcentral/fileexchange/8277. 2012.

[53] C. Blivet, J.-F. Larché, Y. Israëli, and P.-O. Bussière, "Non-Arrhenius behavior: Influence of the crystallinity on lifetime predictions of polymer materials used in the cable and wire industries," *Polym. Degrad. Stab.*, vol. 199, pp. 109890. 2022.

**Shi-Shun Chen** received the Bachelor's degree in safety engineering from Beihang University, Beijing, China, in 2021. He is currently pursuing the Ph.D. degree at the School of Reliability and Systems Engineering, Beihang University, Beijing, China. He is also a visiting Ph.D. student with the Energy Department, Politecnico di Milano, Milan, Italy. His research interests include accelerated testing, reliability experiment and meta-heuristic algorithms.

**Dong-Hua Niu** received the Master's degree in physics from Beihang University, Beijing, China, China, in 2017. He is currently the R & D Engineer of Beijing Tianyu Aerospace New Materials Technology Co., Ltd, Beijing, China. His research interests include the development of thermal control materials for spacecraft.

**Wen-Bin Chen** received the Ph.D. degree in Systems Engineering from Beihang University, Beijing, China, in 2022. He is currently a postdoctoral researcher at the School of Reliability and Systems Engineering, Beihang University, Beijing, China, and a youth member in the sixth council of the Uncertain System Division of the Operations Research Society of China. His research interests include system reliability modeling and reliability experiment.

**Jia-Yun Song** received the Master's degree in mechanical electronics engineering from Beijing Institute of Technology, Beijing, China, in 2010. He is currently the vice general manager of Beijing Tianyu Aerospace New Materials Technology Co., Ltd, Beijing, China. His research interests include the measure detection and the aerospace technology application industry.

**Ya-Fei Zhang** received the Master's degree in material science and engineering from University of Science and Technology Beijing, Beijing, China, in 2008. He is currently the R & D Engineer of Beijing Tianyu Aerospace New Materials Technology Co., Ltd, Beijing, China. His research interests include the development of thermal control materials for spacecraft.

**Xiao-Yang Li** received the Ph.D. degree in aerospace systems engineering from Beihang University, Beijing, China, in 2007. She is currently a Professor, and the director of the Department of Systems Engineering with the School of Reliability and Systems Engineering, Beihang University, Beijing, China. Her research interests include accelerated testing, system reliability modeling, and reliability experiment.

**Enrico Zio** received the Ph.D. degree in nuclear engineering from Politecnico di Milano, in 1996, and also from Massachusetts Institute of Technology, Cambridge, MA, USA, in 1998. He is currently the Director of the Chair on Systems Science and the Energetic Challenge of the Foundation, Électricité de France, Paris, France, CentraleSupélec, Gif-sur-Yvette, France, and a Full Professor and the President of the Alumni Association, Politecnico di Milano. His research interests include the modeling of the failure-repair-maintenance behavior of components and complex systems, for the modeling and analysis of their reliability, maintainability, safety, vulnerability, and resilience characteristics, by Monte Carlo simulation methods, artificial intelligence techniques, and optimization heuristics.